\documentclass[apj]{emulateapj}

\usepackage{graphicx}
\usepackage{amsmath}
\usepackage{ulem}
\usepackage{cases}

\usepackage{color}\def\red#1{\textcolor{red}{#1}}

\usepackage{natbib}

\newcommand{\feoh}{[{\rm Fe} / {\rm H}]}
\newcommand{\msun}{\, M_\odot}
\newcommand{\Zsun}{\, Z_\odot}

\newcommand{\mmd}{M_{\rm md}}

\def\abra#1#2{[{\rm #1}/{\rm #2}]}
\def\bfrac#1#2{\left( \frac{#1}{#2} \right)}
\def\Y#1{Y_{\rm #1}}

\shorttitle{R-process chemical evolution}
\shortauthors{Komiya \& Shigeyama}

\begin{document}
\title{Contribution of Neutron Star Mergers to the R-process Chemical Evolution in the Hierarchical Galaxy Formation}

\author{Yutaka Komiya\altaffilmark{1} and Toshikazu Shigeyama\altaffilmark{1}}
\altaffiltext{1}{Research Center for the Early Universe, University of Tokyo, Hongo 7-3-1, Bunkyo-ku, 113-0033, Tokyo, Japan}

\begin{abstract}
The main astronomical source of r-process elements has not yet been identified. 
One plausible site is neutron star mergers (NSMs), 
 but from perspective of the Galactic chemical evolution, 
 it has been pointed out that 
NSMs cannot reproduce the observed r-process abundance distribution of metal-poor stars at $\feoh < -3$. 
Recently, Tsujimoto \& Shigeyama (2014) pointed out that NSM ejecta can spread into much larger volume than ejecta from a supernova. 
We re-examine the enrichment of r-process elements by NSMs considering this difference in propagation using the chemical evolution model under the hierarchical galaxy formation.  
 The observed r-process enhanced stars around $\feoh \sim -3$ are reproduced if the star formation efficiency is lower for low-mass galaxies under a realistic delay time distribution for NSMs.  
We show that a significant fraction of NSM ejecta escape from its host proto-galaxy to pollute intergalactic matter and other proto-galaxies. 
The propagation of r-process elements over proto-galaxies changes the abundance distribution at $\feoh < -3$ and obtains distribution compatible with observations of the Milky Way halo stars. 
In particular, the pre-enrichment of intergalactic medium explains the observed scarcity of EMP stars without Ba and abundance distribution of r-process elements at $\feoh \lesssim -3.5$.  
\end{abstract}

\section{Introduction}
The rapid neutron capture process (r-process) is one of main processes to synthesize elements heavier than the iron group. 
The dominant r-process element source in the universe is still matter of debate. 
Two possible sites have been proposed; core-collapse supernovae \citep[CCSNe, e.g.,][]{Sato74, Woosley92}, and coalescence of binaries of double neutron star (NS-NS) and black hole-neutron star (BH-NS) systems \citep[e.g.,][]{Lattimer74, Rosswog99}. 
We refer to scenarios based on chemical evolution models of r-process elements exclusively supplied from binary neutron star mergers (NSMs) as the NSM scenarios in the following.

One promising way to understand the enrichment history of heavy elements in the universe is observations of extremely metal-poor (EMP) stars.  
These stars are the living fossils born in the early universe and can be probes for the Galactic chemical evolution. 
Europium (Eu) is commonly used as an element representative of r-process elements but 
we also discuss with barium (Ba) abundance as another representative element since it is more easily observed in EMP stars than Eu. 
Though Ba in the solar system is mainly produced by s-process in intermediate mass stars on the asymptotic giant branch (AGB), 
the contribution from s-process is thought to be negligible for EMP stars since their formation epoch is in prior to the period when the first intermediate mass stars in the universe evolve to AGB stars. 
We note that the measured abundance ratios between Eu and Ba in EMP stars are similar to the ratio of the r-process components of these elements in the solar system, except for carbon-enhanced EMP stars, of which the surface abundances have probably been changed by binary mass transfer.

It is known that the r-process abundances of EMP stars with $\feoh \lesssim -2.5$ show a very large star-to-star scatter of $\sim 3 \ {\rm dex}$ \citep[e.g.,][]{Burris00, Honda04}. 
The average abundance of Ba shows a decreasing trend as metallicity decreases in the metallicity range of $-3.3 < \feoh < -2.6$, while it seems to reach a plateau at $\feoh < -3.3$ \citep{Andrievsky09}. 
Based on this trend, \citet{Roederer13} and \citet{Komiya14} pointed out that we can detect Ba on almost all the EMP stars.

It has been pointed out from the viewpoint of chemical evolution that r-process elements originating from neutron star mergers (NSMs) have difficulties in realizing the observed abundance features in EMP stars.
\citet{Argast04} argued that the model with NSM as the dominant r-process source predicts too large abundance scatter since NSMs have a very low event rate \citep[$ 1 - 300 \ {\rm Myr}^{-1}$ in the Milky Way (MW), ][]{Belczynski02} and very large r-process yield \citep[$\sim 10^{-1} - 10^{-3} \msun$, e.g.,][]{Rosswog99} per event. 
In addition, it has been argued that the long delay-time of NSM ($\gtrsim 10^8$yr) is incompatible with the presence of r-process enhanced stars at $\feoh < -2.5$.  

On the other hand, from other points of view, there is growing evidence that the NSMs are major r-process sources. 
Theoretical studies of nucleosynthesis in CCSNe show difficulties in synthesizing elements of the second and third r-process peaks, in the framework of neutrino driven explosion \citep[e.g.,][]{Wanajo11, Wanajo13}, 
while these elements are successfully synthesized in the NSM ejecta \citep[e.g.,][]{Bauswein13, Wanajo14}. 
Observationally, a short gamma-ray burst GRB~130603B showed enhancement of infrared radiation in its afterglow \citep{Tanvir13, Berger13}. 
This event agrees with the predicted properties of a ``kilonova", radiation from matter heated by radioactive decay of r-process ejecta from a NSM \citep{Tanaka13, Barnes13}. 
\citet{Hotokezaka15} pointed out that the discrepancy in abundances of the radioactive r-process element $^{244}$Pu between the early solar system and the current solar vicinity may indicate a very low event rate for the r-process source. 
The discovery of an r-process rich ultra-faint dwarf galaxy \citep{Ji16, Roederer16} also favors the r-process source with a large yield and rare event rate.

Recently, \citet{Tsujimoto14} pointed out that the ejecta from a NSM can spread into very large volume since the NSM ejecta expand at very large velocities of $10 - 30\%$ of the speed of light \citep[e.g.,][]{Goriely11, Hotokezaka13}. 
The stopping length, $l_{\rm s}$, of $^{153}$Eu with speed of 0.2$c$ is $l_{\rm s} \sim2.6 (n/1 {\rm cm}^{-3})^{-1}$kpc, which can be larger than the radius of a dwarf spheroidal galaxy. 
(See Section 2.3.1 for the derivation of $l_{\rm s}$. Note that \citet{Tsujimoto14} evaluate this value as $\sim 400 \ {\rm kpc}$, which is in error by a factor of 153.)
We investigate effects of this large-scale spreading of the NSM ejecta on the Galactic chemical evolution.

Recent chemical evolution studies confirmed that the emergence of Eu around $\feoh \sim -3$ cannot be explained by the NSM scenario, 
unless binary NSs coalesce on very short timescales of 10 Myr or less \citep{Matteucci14, Cescutti15, Wehmeyer15}, 
while both observations of binary pulsars \citep{Lorimer08} and models of binary population synthesis \citep{Dominik12, Kinugawa14} indicate average coalescence timescales beyond 100 Myr. 
In addition, even with the short delay time, the secondary r-process source such as CCSNe is invoked to reproduce the abundance distribution at $\feoh < -3$ \citep{Cescutti15, Wehmeyer15}. 
\citet{Ishimaru15} show that the appearance of r-process elements at $\feoh \sim -3$ can be explained by the combination with the coalescence timescale of 100 Myr for the majority (95\%) of NSMs and 1 Myr for the rest (5\%), 
if the star formation efficiencies are lower for less-massive sub-halos. 
But they use a one-zone model and do not account for the scatter of $\abra{r}{Fe}$. 
\citet{Hirai15} follow the enrichment history of r-process elements by chemo-dynamical simulations but only for a dwarf galaxy. 
\citet{vandeVoort15} and \citet{Shen15} compute the formation of the MW by chemo-dynamical simulations
but their models lack the resolution to correctly follow the formation of Pop III and early Pop II stars.  
We note that the results of these simulation studies are dependent on assumptions about the sub-grid mixing process and numerical resolution.

All these previous studies do not take into account the difference of the propagation process between NSM ejecta and SN ejecta. 
In our previous study, we investigated the enrichment of r-process elements using a chemical evolution model with merging history of proto-galaxies, 
 but without considering a difference in propagation \citep{Komiya14}. 
We pointed out that the observed very low frequency of Ba lacking stars is in contradiction with the NSM scenario even when we assume a significant surface pollution. 
In this paper, we revisit this problem by considering the possibility that NSM ejecta can escape from its host proto-galaxy and pollute neighboring proto-galaxies and intergalactic medium (IGM).

The structure of this paper is as follows. 
Next section describes our model and its ingredients. 
Section~\ref{resultS} presents results and compares them with observations. 
Section~\ref{summaryS} summarizes the paper.

\section{Model description}\label{modelS}

We use the hierarchical chemical evolution model presented in \citet{Komiya14, Komiya15}. 
In the following, we give a brief summary of the model and changes from the previous works.

\subsection{The hierarchical chemical evolution model}
Our model treats individual metal poor stars in the framework of the hierarchical galaxy formation. 
This model also considers gas infall and outflow, metal pre-enrichment in IGM, and surface pollution of stars by accretion of interstellar matter (ISM).

\subsubsection{Merger Tree}
We build merger trees of dark-matter halos based on the extended Press-Schechter theory \citep{SK99}. 
The first stars are formed in halos with $T_{\rm vir} > 10^3 {\rm K}$, 
 where $T_{\rm vir}$ is the virial temperature, and $T_{\rm vir}=10^3 {\rm K}$ corresponds to the halo mass of $M_{\rm h} = 1.9\times 10^5 \msun$ at a redshift of $z = 30$. 
We refer to the baryonic component in the branches of the merger tree as proto-galaxies, which are low-mass galaxies as building blocks of the Milky Way.

Each halo hosts one proto-galaxy at the beginning. 
We assume that when two dark halos merge, the proto-galaxy in the more massive halo is placed at the center of the newly formed halo and the proto-galaxy in the smaller halo becomes a satellite in a sub-halo and that the mass of the smaller halo remains unchanged. 
We neglect tidal stripping and ram pressure stripping. 
The orbital energy of the satellite decays through dynamical friction \citep{LC93}. 
The satellite merges with the central galaxy on the dynamical friction timescale, $\tau_{\rm df}$, given by the standard Chandrasekhar formula, and stars in the satellite are added to the stellar halo of the central galaxy \citep[e.g.,][]{Cole00}.

\subsubsection{Star Formation, Metal Enrichment and Gas Outflow}
The star formation rate, $\dot{M_*} = \epsilon_* M_{\rm gas}$, is expressed as the product of the gas mass, $M_{\rm gas}$, and the star formation efficiency (SFE), $\epsilon_*$. 
The masses of individual stars are set randomly following the initial mass function (IMF) described in section~\ref{S:IMF}.

Metal yields of supernovae (SNe) are adopted from theoretical results of \citet{Kobayashi06} for type II SNe and \citet{Nomoto84} for type Ia SNe. 
We assume homogeneous mixing of metal in each proto-galaxy.

We consider gas outflow triggered by individual SNe. 
The first SN in a proto-galaxy form a galactic wind, and subsequent SNe add gas and energy to the wind. 
For the outflow energy, $E_{\rm w}$, we use the following formula as a function of the SN kinetic energy, $E_{\rm k}$, and the gas binding energy, $E_{\rm bin}$, of a proto-galaxy, 
\begin{equation}
E_{\rm w} = E_{\rm k} \frac{\epsilon_{\rm o} +E_{\rm k}/E_{\rm bin}}{1+E_{\rm k}/E_{\rm bin}},
\end{equation}
which supposes that the outflow carries away all the SN kinetic energy if $E_{\rm k} \gg E_{\rm bin}$ but a fraction, $\epsilon_{\rm o}$, of the kinetic energy turns to the outflow if $E_{\rm k} \ll  E_{\rm bin}$. 
We compute inhomogeneous metal enrichment in the IGM by following the evolution of the winds from proto-galaxies assuming the momentum conserving snow-plow shell model.

In the fiducial model, 
 we adopt the star formation efficiency depending on the mass of a halo, $\epsilon_* \propto  M_{\rm h}^{0.3}$,
 motivated by the observational mass-metallicity relation, 
 while a constant efficiency was adopted in our previous studies. 
The observed correlation between stellar masses, $M_*$, and average metallicities, $\left<\feoh\right>$, of galaxies is expressed as $10^{\left<\feoh\right>} \propto  M_*^{0.3}$ \citep{Kirby13}, and  
 may indicate a lower star formation efficiency in a lower mass galaxy \citep{Ishimaru15}. 
 We use $M_{\rm h}^{0.3}$ instead of $M_*^{0.3}$ to avoid $\epsilon_* = 0$ at the beginning. 
Measurements of rotation curves of low mass star forming galaxies showed that the halo mass is almost proportional to the stellar mass \citep{Miller14} .

We also present results of the model with a constant SFE but with a higher outflow rate.  
We adopt $\epsilon_{\rm o} = 0.5$ in the constant SFE model rather than $\epsilon_{\rm o} = 0.1$ as the fiducial value in our previous study. 
This model yields a slope of mass-metallicity relation very similar to the fiducial model with $\epsilon_* \propto M_{\rm h}^{0.3}$ and $\epsilon_{\rm o} = 0.1$.

We evaluate the proportional constant of the SFE to yield the solar metallicity in the current MW,  
$\epsilon_* = 1.2 \times 10^{14} (M_{\rm h}/\msun)^{0.3} {\rm yr}^{-1}$ in the fiducial model and $\epsilon_* = 3.4 \times 10^{11} {\rm yr}^{-1}$ in the constant SFE model.


\subsubsection{Initial Mass Function}\label{S:IMF}
For the initial mass function (IMF), $\xi$, we use the log-normal function with a power-law tail at higher mass part as \citet{Chabrier03} adopted, 
\begin{equation}
\xi(\log m) \propto \begin{cases}
  \exp\left\{ -\frac{(\log m - \log \mmd)^2}{2 \sigma^2} \right\}	& (m \leq m_{\rm norm}) \\
  m^{-1.3}	& (m > m_{\rm norm}).
\end{cases}
\end{equation}
\citet{Chabrier03} yielded $\mmd = 0.22\msun$, $\sigma = 0.33$, and $m_{\rm norm} = 0.7\msun$ for the Galactic spheroid.  
For EMP stars, however, we have shown that the low-mass peaked IMF obtains a much smaller frequency of carbon enhanced stars and more numerous EMP stars than observed by the HES survey \citep{Komiya09, Komiya09L, Komiya15}. 
A high or intermediate-mass peaked IMF shows a better consistency with observations. 
Here, we compute models with the IMFs peaked at $\mmd = 0.22\msun, 2\msun$, and $10\msun$ ($m_{\rm norm}$ is also shifted in proportion to $\mmd$).  
The binary fraction is set to be 0.5, and we adopt the flat mass ratio distribution. 
We note that the predicted r-process abundance distribution is not sensitive to the choice of the peak mass. 
 ( In previous studies, we may have overestimated the volume observed by the HES survey due to the assumption of $L = 100L_\odot$ for giant stars and applying an incorrect limiting magnitude.  The correction slightly lower the estimated value of $\mmd$ but it still remains in the intermediate mass range. )

\subsubsection{Surface Pollution of Stars}
We consider that the surface of a star is polluted by accreting ISM.
We trace changes of surface abundance of stars with $\feoh < -2.5$ using the Bondi accretion formula. 
The surface pollution can be significant on the surface abundances of EMP stars, as shown in \citet{Komiya14, Komiya15}.

\subsubsection{Enrichment History of Fe and Mg}
\begin{figure}
\includegraphics[width=0.7\columnwidth]{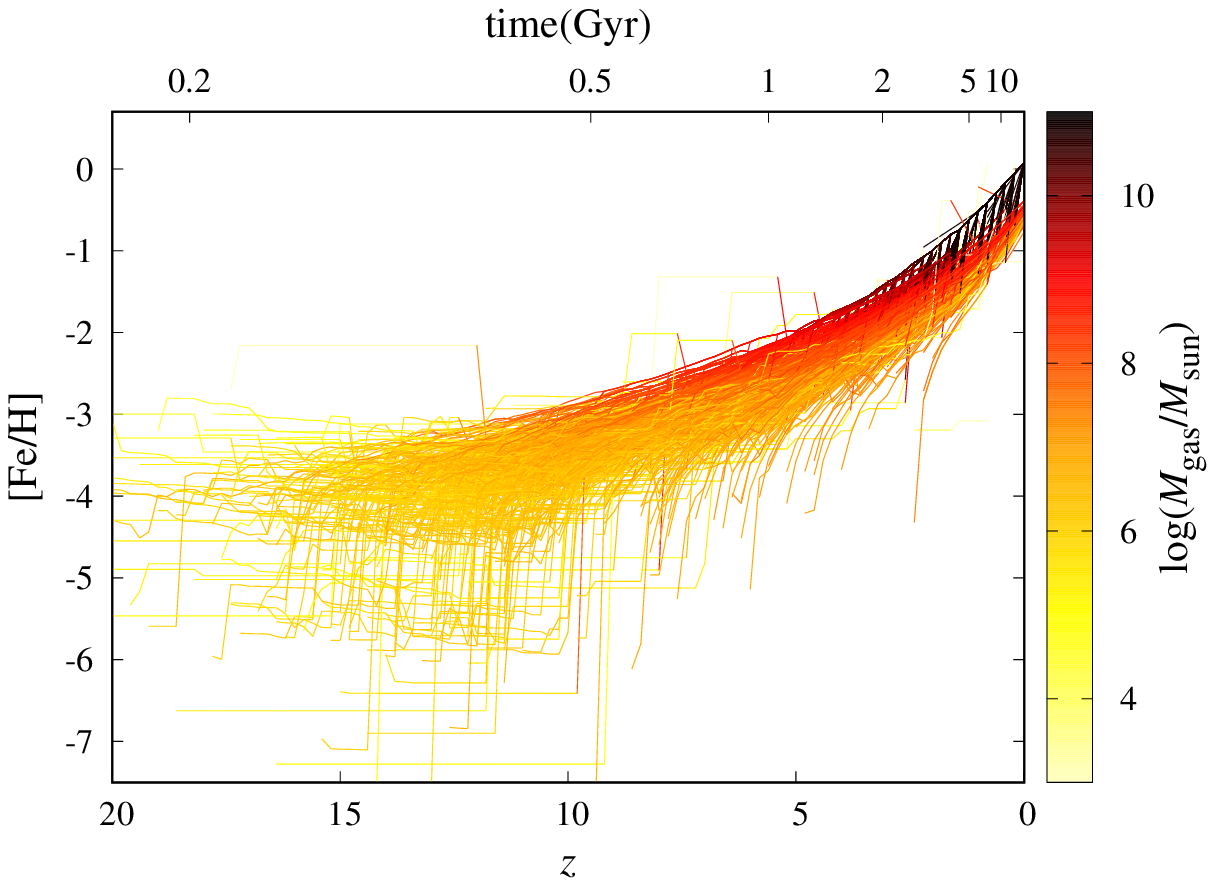} \\
\includegraphics[width=0.7\columnwidth]{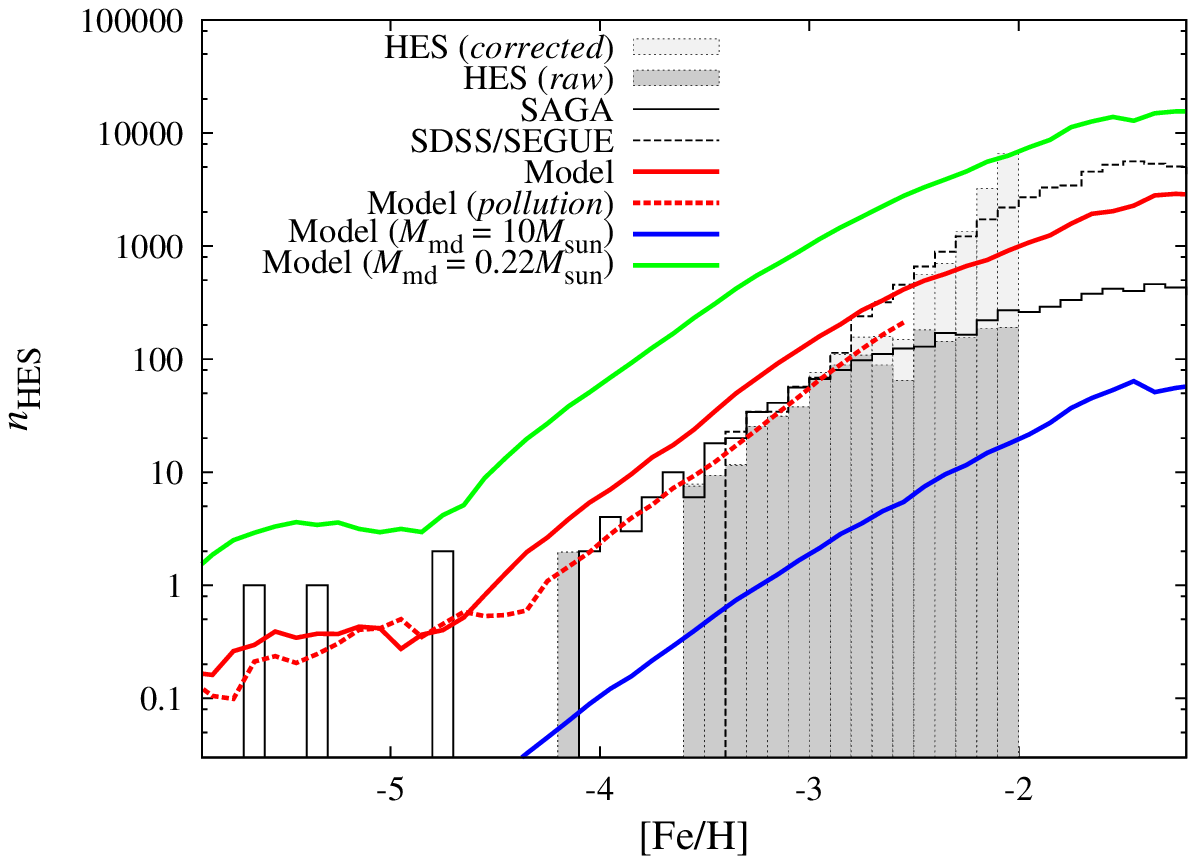} \\
\caption{ 
Top panel: 
$\feoh$ of proto-galaxies against redshift, with the gas mass of proto-galaxies color coded. 
Bottom panel:
Metallicity distribution functions (MDFs) of metal-poor stars in the MW halo. 
The green, red, and blue lines show the predicted number of giant stars in the survey volume of the HES survey for models with $\mmd = 0.22\msun, 2\msun, 10\msun$, respectively. 
The dashed red line shows the MDF after surface pollution. 
The shaded histogram is the raw (dark-gray) and bias-corrected (light-gray) MDFs by the HES survey \citep{Schorck09}. 
The solid black line is the high-resolution sample biased toward lower metallicity by the SAGA database \citep{Suda08} and the dashed black line is the SDSS/SEGUE sample by \citet{Carollo10}, scaled to fit the HES data. 
The color scale shows the predicted number density and the black crosses are the observational sample from the SAGA database.
}\label{MDF}
\end{figure}

The top panel of Figure~\ref{MDF} shows the metal enrichment history in our fiducial model. 
We plot $\feoh$ of sampled proto-galaxies against redshift.  
Color denotes the gas mass of proto-galaxies. 
In our model, $\feoh$ can decrease when a proto-galaxy merges with a more metal deficient proto-galaxy. 
Most of stars with $\feoh<-3$ are formed at $z>5$, and stars with $-3 < \feoh < -2$ are formed at $ 3 \lesssim z \lesssim 10$.

We show the resultant metallicity distribution functions (MDFs) in the bottom panel.  
The green, red, and blue lines show the predicted MDFs in the cases of $\mmd = 0.22\msun, 2\msun, 10\msun$, respectively. 
The solid and dashed red lines are the intrinsic metallicity and the surface metallicity after pollution, respectively.  
Shaded histograms show the raw (dark-shaded) and bias-corrected (light-shaded) MDFs of the HES sample \citep{Schorck09}. 
We scale the predicted MDFs considering the field of view ($6726 \ {\rm deg}^2$) and the magnitude limit ($B = 17.2$) of the HES survey \citep{Christlieb08}.   
Since most of the sample stars in \citet{Schorck09} are around the base of the red giant branch, we count the predicted number of giant stars in the survey volume assuming that the typical luminosity is $L = 20L_\odot$. 
We employ the de Vaucouleur density profile for the MW halo. 
We also show the MDF of the SDSS/SEGUE sample \citep[][dashed black line]{Carollo10} scaled to match the HES data and the MDF by the SAGA database (solid black line), which collects stellar abundances by high-resolution spectroscopy of metal-poor stars from literatures \citep{Suda08}.

As seen in the bottom panel, 
though all the three IMFs lead to a similar form of the MDF, the IMF with $\mmd = 2\msun$ solely reproduces the total number. 
The IMF with a higher $\mmd$ produces a fewer number of low-mass stars which can survive to date,  
and at the same time form more numerous massive stars producing iron to shorten the timescale for low-metallicity star formation.   
We show results of the models with  $\mmd = 2\msun$ in section~\ref{resultS}.  
The main results about r-process elements remain true for the other two $\mmd$.  
The MDF with $\mmd = 2\msun$ slightly underestimates the number of stars at $\feoh \gtrsim -2.5$. 
It may be due to the change of the IMF to the low-mass peaked one \citep{Suda13, Yamada13}.

\begin{figure}
\includegraphics[width=1\columnwidth]{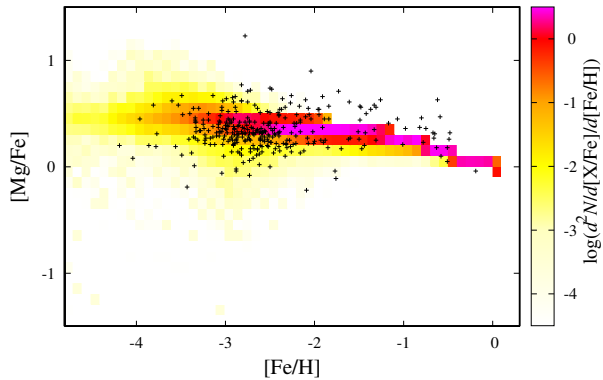}
\caption{ 
Distribution of abundance ratios of magnesium to iron against metallicity. 
The color scale shows the predicted number density and the black crosses are the observational sample from the SAGA database.
}\label{MgFe}
\end{figure}

Our model obtains the abundance distribution of alpha elements consistent with observations. 
We present the correlation of the abundance ratios of magnesium to iron with metallicity in Figure~\ref{MgFe}, and oxygen and silicon shows the similar abundance distributions. 

\subsection{Neutron Star Merger}

The mass range for progenitor stars of NSs is $8 - 25 \msun$. 
We assume that $1\%$ of binary systems in this mass range form NS binaries which coalesce within the age of the universe. 
This yields one coalescing NS binary out of 1000 massive stars.  
We adopt the delay time distribution of the NSM, 
\begin{equation}
\frac{d n_{\rm NSM}}{d t_{\rm d}} \propto t_{\rm d}^{-1}
\end{equation}
 for $10^7 {\rm yr} < t_{\rm d} < 10^{10} {\rm yr}$ following results of binary population synthesis by \citet{Dominik12}. 
In this case, the averaged delay time is $\left< t_{\rm d} \right> \sim 1.3 \times 10^9 {\rm yr}$. 
The NSM event rate is proportional to the following integration,  
\begin{equation}
 \int^{10^{10} {\rm yr}}_{10^{7} {\rm yr}} \dot{M_*}(t-t_{\rm d}) \frac{d n_{\rm NSM}}{d t_{\rm d}}(t_{\rm d}) dt_{\rm d}. 
\end{equation}

The ejecta mass of a NSM is assumed to be $Y_{r} = 0.03\msun$, based on the estimation of ejecta mass of the observed kilonova candidate GRB 130603B \citep{Tanvir13, Berger13}. 
We assume the scaled solar r-process abundance pattern \citep{Arlandini99} for NSM ejecta. 
The elemental yields of Ba and Eu are $\Y{Ba}=1.3\times 10^{-3}\msun$ and $\Y{Eu}=1.5\times 10^{-4}\msun$, respectively.

We do not distinguish NS-NS mergers and BH-NS mergers in our computation, while recent numerical studies show that a BH-NS merger can eject lager mass than a NS-NS merger \citep{Hotokezaka13}. 
In this paper, we do not consider other heavy element sources such as s-process in intermediate massive stars or r-process in SNe~II.

\subsection{Propagation of NSM ejecta}

A significant fraction of the NSM ejecta can escape from proto-galaxies. 
Nearby proto-galaxies can capture the escaped r-process elements. 
The rest of r-process ejecta travel away into the intergalactic space with high velocities. 
The r-process elements cool and are coupled with IGM, and are incorporated into galaxies along gas infall as described in the following.

\subsubsection{Escape from a Proto-Galaxy}\label{S:fesc}

R-process elements from a NSM propagate until they lose kinetic energies by interaction with ISM.
The dominant energy deposition process is the ionization of neutral hydrogen in HI gas, and the Coulomb scattering in the ionized gas. 
The energy loss rate is 
\begin{eqnarray}\label{eq:omeganeutral}
\omega \equiv  &-\frac{dE}{dt} = 1.82 \times 10^{-7} Z^2 [1 - 1.034 \exp (-137\beta & Z^{-0.688})]  \nonumber \\
&\times n_{\rm HI} (1+0.0185 \ln \beta) \beta^{-1} \hbox{ eV s}^{-1}	 
\end{eqnarray} 
in the neutral gas and 
\begin{equation}\label{eq:omegaplasma}
\omega = 3.1 \times 10^{-7} Z^2 n_e \beta^{-1} \hbox{ eV s}^{-1}   
\end{equation} 
in the ionized plasma \citep{Schlickeiser02}, 
where $n_{\rm HI}$ and $n_e$ is the number density of neutral hydrogen and free electron, respectively, $\beta$ is the velocity in units of the speed of light and set at $\beta \equiv v_r/c = 0.2$ in this study, and $Z$ is the atomic number. 
The stopping length, 
\begin{equation}\label{eq:ls}
l_{\rm s} \equiv \frac{(\gamma-1) m_{\rm p} A c^2 }{\omega} v_r \sim 2.6 \bfrac{n_{\rm HI}}{1 \ {\rm cm}^{-3}}^{-1} \ {\rm kpc}
\end{equation}
 for $^{153}{\rm Eu}$ in the neutral gas, where $m_{\rm p}$ is the proton mass, $A$ is the mass number, and $\gamma$ is the Lorentz factor.

We assume that r-process elements ejected from an NSM travel the stopping length of $l_{\rm s}$ on a straight line at a constant velocity.  
The ejected particles stop rather suddenly once they travels the stopping length since the energy loss rate increases as the energy decreases. 
The fraction, $f_{\rm esc}$, of NSM ejecta escaping from the proto-galaxy is estimated as
\begin{equation}\label{eq:fesc}
 f_{\rm esc} = \exp\left(-\frac{R_{\rm g}}{l_{\rm s}}\right) ,
\end{equation} 
where $ R_{\rm g}$ is the radius of the host proto-galaxy. 
For simplicity, we assume that a proto-galaxy is a homogeneous sphere of neutral gas, and an NSM takes place at the center of the proto-galaxy.

We give the gas density and radius of proto-galaxies assuming isobaric cooling of the virialized gas \citep{Komiya15}, 
 i.e., $n = \rho_{\rm vir} (\mu m_{\rm p})^{-1} \left({M_{\rm gas}}/{M_{\rm h}}\right) \left({T_{\rm gas}}/{T_{\rm vir}}\right)^{-1}$,
 and 
 $ R_{\rm g} = R_{\rm vir} (T_{\rm gas}/T_{\rm vir})^{1/3} $, 
 where 
 $\rho_{\rm vir}$ is the averaged density of the virialized dark-matter halos as a function of redshift, 
 $\mu$ is the mean molecular weight, 
 $R_{\rm vir}$ is the virial radius of the host halo.  
Gas can be cooled to $T_{\rm cool} \sim 10$ K in the MW but $T_{\rm cool} \sim 200$ K in the primordial mini-halo, and $T_{\rm cool} \sim 50$ K in relic HII regions without metal. 
The gas temperature, $T_{\rm gas}$, is set at the maximum of the temperature of the cosmic microwave background or $T_{\rm cool}$. 
These assumptions yield $n = 4.2 \ {\rm cm}^{-3}$ and $R_{\rm g} = 70$ pc for the primordial proto-galaxy with $M_{\rm h} = 10^{6}\msun$ at $z=20$, 
 and $n = 3.9 \ {\rm cm}^{-3}$ and $R_{\rm g}=5.2$ kpc for the current MW ($z=0, M_{\rm h} = 10^{12}\msun$).

\subsubsection{Capture by Other Proto-galaxies}
In this study, we assume isotropic mass ejection from a NSM \citep{Hotokezaka13}. 
The fraction of escaped NSM ejecta encountering with a proto-galaxy with radius $R_{\rm g}$ is $ {\pi R_{\rm g}^2}/{4 \pi d_{}^2}$, 
 where $d_{}$ is the distance to the proto-galaxy.

For a particle which runs into a spherical proto-galaxy with an impact parameter of $r$, 
the length of the pass crossing the proto-galaxy is $2 \sqrt{R_{\rm g}^2 - r^2}$. 
Under the same assumptions as in eq.~(\ref{eq:fesc}), the fraction of particles that are not captured by the proto-galaxy is $\exp\left(-{2 \sqrt{R_{\rm g}^2 - r^2}}/{l_{\rm s}}\right)$. 
The fraction of captured high-energy r-process elements out of all particles encountering the proto-galaxy is 
\begin{eqnarray}
 f_{\rm cap} & = & \frac{1}{\pi R_{\rm g}^2} \int^{R_{\rm g}}_{0} \left( 1 - \exp\left(-\frac{2 \sqrt{R_{{\rm g}}^2-r^2}}{l_{\rm s}} \right) \right) 2 \pi r dr  \nonumber \\
& = & 1- \frac{1}{2} \left( \frac{l_{\rm s}}{R_{\rm g}}\right) ^2 \left\{ 1- \left(1 + \frac{2 R_{\rm g}}{l_{\rm s}} \right)\exp\left(-\frac{2 R_{\rm g}}{l_{\rm s}}\right) \right\}.  
\end{eqnarray} 
We mention that, in the limit of $R_{\rm g} \ll l_{\rm s}$, the capture rate approaches to $ f_{\rm cap} \sim \frac{4}{3} R_{\rm g}/l_{\rm s} $ and the total amount of captured particles is proportional to $ {\pi R_{\rm g}^2}f_{\rm cap} \sim \frac{4}{3} \pi R_{\rm g}^3/l_{\rm s} \propto M_{\rm gas}$ since $l_{\rm s} \propto n^{-1}$ (eq.~\ref{eq:ls}), i.e., the capture rate is simply proportional to the gas mass but independent of radius and density.

We estimate the distance between the $n$-th and the $m$-th proto-galaxies using the following equation, 
\begin{subnumcases}
{d_{n,m}(t) = }
  r(t|M_{{\rm t}, n,m}, t_{{\rm c}, n,m}) 	\label{eq:dhalo} \\
   \qquad \qquad \qquad \qquad \qquad \qquad (t < t_{{\rm c}, n,m}) \nonumber \\
  d_{{\rm sat},m}(t) =  R_{{\rm vir},n}  \left( 1-\frac{t- t_{{\rm c},n,m}}{\tau _{\rm df}} \right) \label{eq:dsat}  \\
  \qquad \qquad \qquad  (t_{{\rm c}, n,m} \leq t < t_{{\rm c}, n,m} + \tau _{\rm df})  \nonumber \\
  \sqrt{ d_{{\rm sat}, n} + d_{{\rm sat}, m} - 2 d_{{\rm sat},n} d_{{\rm sat},m} \cos\theta }   \label{eq:d2sat} \\
  \qquad \qquad \qquad \qquad \qquad  \text{ (two satellites).} \nonumber 
\end{subnumcases}
In the case of two proto-galaxies in different dark halos (eq.~\ref{eq:dhalo}), 
we estimate the distance between them under the spherical collapse approximation of halos following \citet{Komiya16}. 
Here two halos are supposed to merge at time $t_{{\rm c}, m,n}$ to become a single halo with the mass of $M_{{\rm t}, m,n}$, and  $r$ is the distance between the two halos (see section 4.1 in \citet{Komiya16} for details). 
After the two dark halos merge (eq.~\ref{eq:dsat}), the proto-galaxy in the more massive halo (labeled $n$) is placed at the center of the merged halo and the other one (labeled $m$) becomes a satellite. 
The satellite galaxy is assumed to approach the center as the distance, $d_{{\rm sat},m}(t)$, from the center linearly decreases as a function of time.
$\tau _{\rm df} $ is the merger timescale by the dynamical friction and we use the formula given in \cite{Cole00}. 
At $t = t_{{\rm c}, m,n} + \tau _{\rm df}$, the satellite merges to the central galaxy. 
In the case of two satellites in the same dark halo (eq.~\ref{eq:d2sat}), we estimate the distance assuming the random distribution for the orbital angles, $\theta $, between the satellite galaxies.

\subsubsection{In the Intergalactic Space}
The energy loss timescale, $\tau_s \equiv (\gamma-1) m_{\rm p} A c^2/\omega$, of $^{153}$Eu propagating in the IGM is given as 
\begin{eqnarray}
\tau_{s, \rm IGM} &=&
\begin{cases}
 430 \bfrac{n_{\rm IGM}}{10^{-4} {\rm cm}^{-3}}^{-1} {\rm Myr}	& {\text {\text (before reionization)}}	\\
 31  \bfrac{n_{\rm IGM}}{10^{-4} {\rm cm}^{-3}}^{-1} {\rm Myr}	& {\text {\text (after reionization)}}	
\end{cases}
\end{eqnarray} 
where we use the energy loss rate for neutral gas (eq.~\ref{eq:omeganeutral}) before reionization and one for ionized gas (eq.~\ref{eq:omegaplasma}) after that. 
We assume the reionization redshift $z = 10$. 
Since the stopping length in the IGM, $l_{s, \rm IGM} \equiv v_r \tau_{s, \rm IGM}$, is larger than the computation box, 
we assume homogeneous mixing of the escaped r-process elements in the intergalactic space, 
while we follow the inhomogeneous metal pollution of the IGM for elements provided by SNe as mentioned above.

R-process elements having escaped from the host galaxy lose their kinetic energies on the timescale of $\tau_{\rm s}$ unless they are captured by another proto-galaxy. 
Such r-process elements are eventually settled in the IGM and some of them are attracted by the gravity of proto-galaxies and accreted onto them along with gas infall.

We can describe the evolution of mass $M_{i, {\rm HE}}$ of the high-energy particle in the intergalactic space and the abundance, $X_{i, \rm IGM}$, of element $i$ in the IGM as follows, 
\begin{align} \label{eq:HE}
\frac{dM_{i,{\rm HE}}}{dt} = 
Y_i \sum_{n} \sum_{k} & \delta(t - t_{{\rm NSM}, n, k}) f_{{\rm esc}, n, i} \left( \sum_{m \neq n} \frac{\pi R_{{\rm g}, m}^2 (1-f_{{\rm cap}, m, i})}{4 \pi d_{n,m}^2} \right) \notag \\
&- \sum_{n} \frac{M_{i,{\rm HE}} \pi R_{{\rm g}, n}^2 v_r}{V_{\rm MW}} f_{{\rm cap}, n, i} - \frac{M_{i,{\rm HE}}}{\tau_{s,{\rm IGM},i}}, 
\end{align} 
\begin{equation}
\frac{d M_{\rm IGM} X_{i, \rm IGM}}{dt} = 
\frac{M_{i,{\rm HE}}}{\tau_{s,i}}
 - \sum_{n} \dot{M}_{{\rm acc},n} X_{i, \rm IGM}, 
\end{equation} 
where $Y_{i}$ is the elemental yields from one NSM event, 
 $t_{{\rm NSM}, n, k}$ is the time when the $k$-th NSM event takes place in the $n$-th proto-galaxy. 
The second term in the right hand side of eq.~(\ref{eq:HE}) describes capture of particles incoming from outside of the computation box, and $V_{\rm MW}$ is the volume of the box, which is a comoving sphere enclosing the total mass of $10^{12}\msun$. 
$\dot{M}_{{\rm acc},n}$ is the gas accretion rate onto the $n$-th proto-galaxy.

Here, we assume that the escaping of the high energy r-process elements from the simulation box balances the incoming from galaxies outside the box. 
We discuss the validity of this assumption in Appendix.

\subsubsection{Evolution of the R-process Abundances}
In summary, evolution of r-process abundances of a proto-galaxy is described as follows, 
\begin{align}
\frac{dM_{{\rm gas}, n} X_{i, n}}{dt} = Y_i \sum_{k} \delta(t - t_{{\rm NSM}, n, k}) (1-f_{{\rm esc},n,i})  \notag \\
 + Y_i \sum_{m \neq n}\sum_{k} \delta(t - t_{{\rm NSM}, m, k}) f_{{\rm esc},m,i}\frac{\pi R_{{\rm g}, n}^2}{4 \pi d_{n,m}^2} f_{{\rm cap}, n,i} \notag \\
 + \frac{M_{i,{\rm HE}} \pi R_{{\rm g},n}^2 v_r}{V_{\rm MW}}   f_{{\rm cap},n,i}
 + \dot{M}_{{\rm acc},n} X_{i, \rm IGM}
\end{align} 
where the first term on the right-hand side is the contribution from NSMs in the proto-galaxy, 
 the second term is the increment by capturing ejecta from NSM events in other proto-galaxies in the simulation box, 
 the third term is high-energy r-process background from galaxies outside the computation box, 
 and the fourth term is the infall of IGM.

\section{Results and Discussion}\label{resultS}

\begin{figure}[h]
\includegraphics[width=\columnwidth]{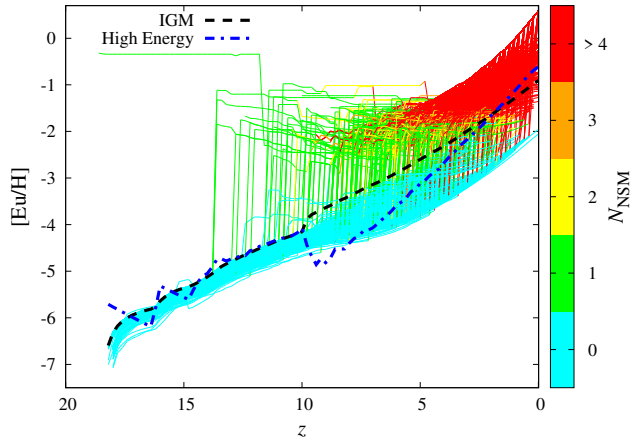}
\caption{
The enrichment history of Eu. 
The thin solid lines show $\abra{Eu}{H}$ of proto-galaxies, with the number of NSM events in them color coded. 
We randomly sample $\sim 1/300$ of proto-galaxies and plot their abundances.  
The dash-dotted blue line denote\red{s} the abundance of high-energy Eu in intergalactic space, $\abra{Eu}{H}_{\rm HE}$, and the dashed black line is the low-energy Eu abundance in IGM.  
}\label{ce}
\end{figure}

Figure~\ref{ce} shows the enrichment history of Eu abundances for sampled proto-galaxies and the IGM (dashed black line). 
The number of NSM events in each proto-galaxy is color coded. 
The dash-dotted blue line denotes the abundance of high energy r-process elements in intergalactic space, 
$\abra{Eu}{H}_{\rm HE} \equiv \log(M_{\rm Eu, HE}/M_{\rm IGM}X_{\rm H})-\log(X_{\rm Eu}/X_{\rm H})_{\odot}$.

The first NSM in this computation run takes place at $z = 18.8$. 
Most of the r-process ejecta escape from its host proto-galaxy to enhance $\abra{Eu}{H}_{\rm HE}$ to 
\begin{eqnarray}
\abra{Eu}{H}_{\rm HE} &=& \log \left( \frac{Y_{\rm Eu} f_{\rm esc}}{M_{\rm IGM} X_{\rm H, IGM}} \right)
 - \log \left( \frac{X_{\rm Eu}}{X_{\rm H}} \right)_\odot  \nonumber \\
 = -5.6 +& \log & \left[  \bfrac{f_{\rm esc}}{1} 
\bfrac{M_{\rm IGM}}{M_{\rm MW} \frac{\Omega_{\rm b}}{\Omega_{\rm M}}}^{-1} 
\bfrac{X_{\rm H, IGM}}{0.8}^{-1} \right], 
\end{eqnarray}
where $\left( X_{\rm Eu}/X_{\rm H} \right)_\odot = 4.7\times 10^{-10}$ is the solar Eu abundance \citep{Grevesse96}, 
and $\Omega_{\rm M}$ and $\Omega_{\rm b}$ are densities of matter and baryon relative to the critical density, respectively.  
In the limit of $R_{\rm g} \ll l_{\rm s}$, the capture rate of high energy r-process particles is proportional to the total gas mass and independent of the density as mentioned,
i.e., the capture rate per unit mass is almost the same for all proto-galaxies with $R_{\rm g} \ll l_{\rm s}$ and for the IGM. 
\abra{Eu}{H} of the IGM (black line) and proto-galaxies (cyan lines) increase in a similar way by capturing high-energy r-process elements at $z > 10$. 
At such a high redshift, majority of baryon is still in the IGM, and most of the escaped NSM ejecta are mixed with the IGM on the timescale of $ \tau_{s, \rm IGM} $.

At $z = 10$, $\abra{Eu}{H}_{\rm HE}$ suddenly decreases because the energy loss rate in the IGM increases due to cosmic reionization. 
The capture rate per unit mass by the ionized IGM becomes several tens of times higher than the one by neutral proto-galaxies.  
Therefore, the abundance of proto-galaxies without NSM events (cyan lines) is $\sim 1$ dex or more smaller than the IGM abundance at $z<10$. 
After the reionization, pre-enrichment of IGM is mainly responsible for the Eu abundance of proto-galaxies without NSM events. 
At $z \lesssim 2$, $\abra{Eu}{H}_{\rm HE}$ overwhelms $\abra{Eu}{H}_{\rm IGM}$ because the energy loss timescale in the IGM increases as the IGM density decreases.

A single NSM event enriches its host proto-galaxy with r-process elements to 
\begin{eqnarray}
\abra{Eu}{H} &=& \log \left( \frac{Y_{\rm Eu} (1-f_{\rm esc})}{M_{\rm gas} X_{\rm H}} \right)
 - \log \left( \frac{X_{\rm Eu}}{X_{\rm H}} \right)_\odot 	\nonumber \\
& \sim &  -1.8 + \log \left[ \bfrac{Y_{\rm Eu}}{1.5\times 10^{-4}\msun} \bfrac{R_{\rm g}}{1{\rm kpc}}  \right. \nonumber \\
 & & \left. \bfrac{n}{1 {\rm cm^{-3}}} \bfrac{M_{\rm gas}}{10^7\msun}^{-1}  \right] 	
\end{eqnarray}
in the limit of $f_{\rm esc} \ll 1$ ($R_{\rm g} \ll l_{\rm s}$). 
When we adopt assumptions for $n$ and $R_{\rm g}$ in section~\ref{S:fesc}, 
we can describe this equation as a function of $T_{\rm gas}, M_{\rm h}$ and the cosmological parameters as follows, 
\begin{eqnarray}
&\abra{Eu}{H} \sim  -1.7 + \log \left[ \bfrac{Y_{\rm Eu}}{1.5\times 10^{-4}\msun} \bfrac{T_{\rm gas}}{30{\rm K}}^{-2/3}
\mu^{2/3}  \right. 	\nonumber \\
& \left. \bfrac{M_{\rm h}}{10^8 h^{-1} \msun}^{-2/9}  \left( \frac{\Omega_{\rm M}(0)}{0.3 \Omega_{\rm M}(z)}\frac{\Delta_{\rm c}}{18\pi}\right)^{8/9} \bfrac{1+z}{10}^{8/3} \bfrac{h}{0.7}^{2}
	\right] ,
\end{eqnarray}
where 
$\Delta_{\rm c}$ is the overdensity of the virialized halo relative to the critical density, 
and $h$ is the dimensionless Hubble parameter.   
These proto-galaxies become much more r-process rich than the intergalactic space, 
and the contribution from other proto-galaxies is negligible for them. 
At $z < 5$, the most massive photo-galaxy shows the highest r-process abundance because of a large event rate and a small escape probability.

\begin{figure*}[]
\includegraphics[width=0.5\textwidth]{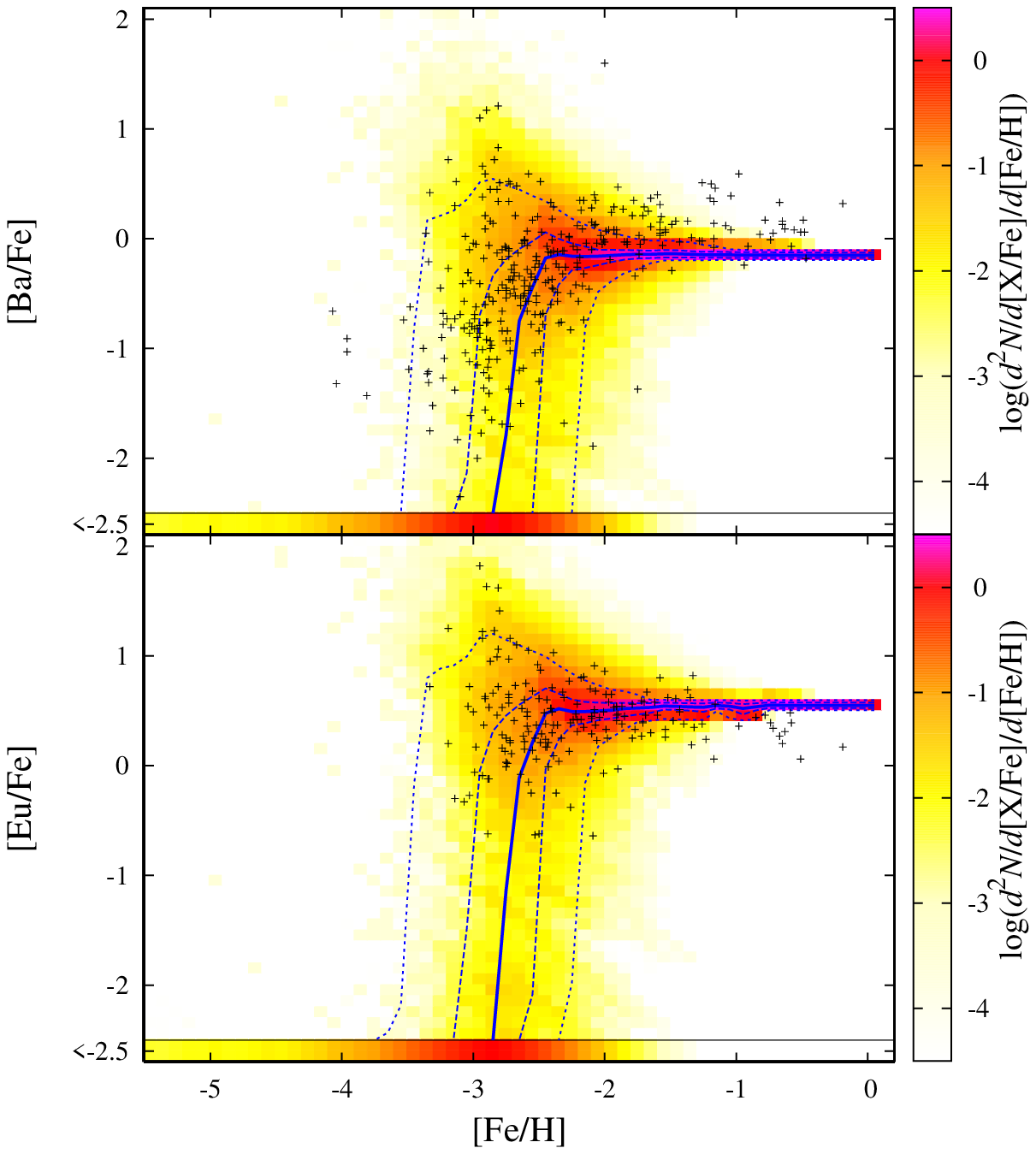}
\includegraphics[width=0.5\textwidth]{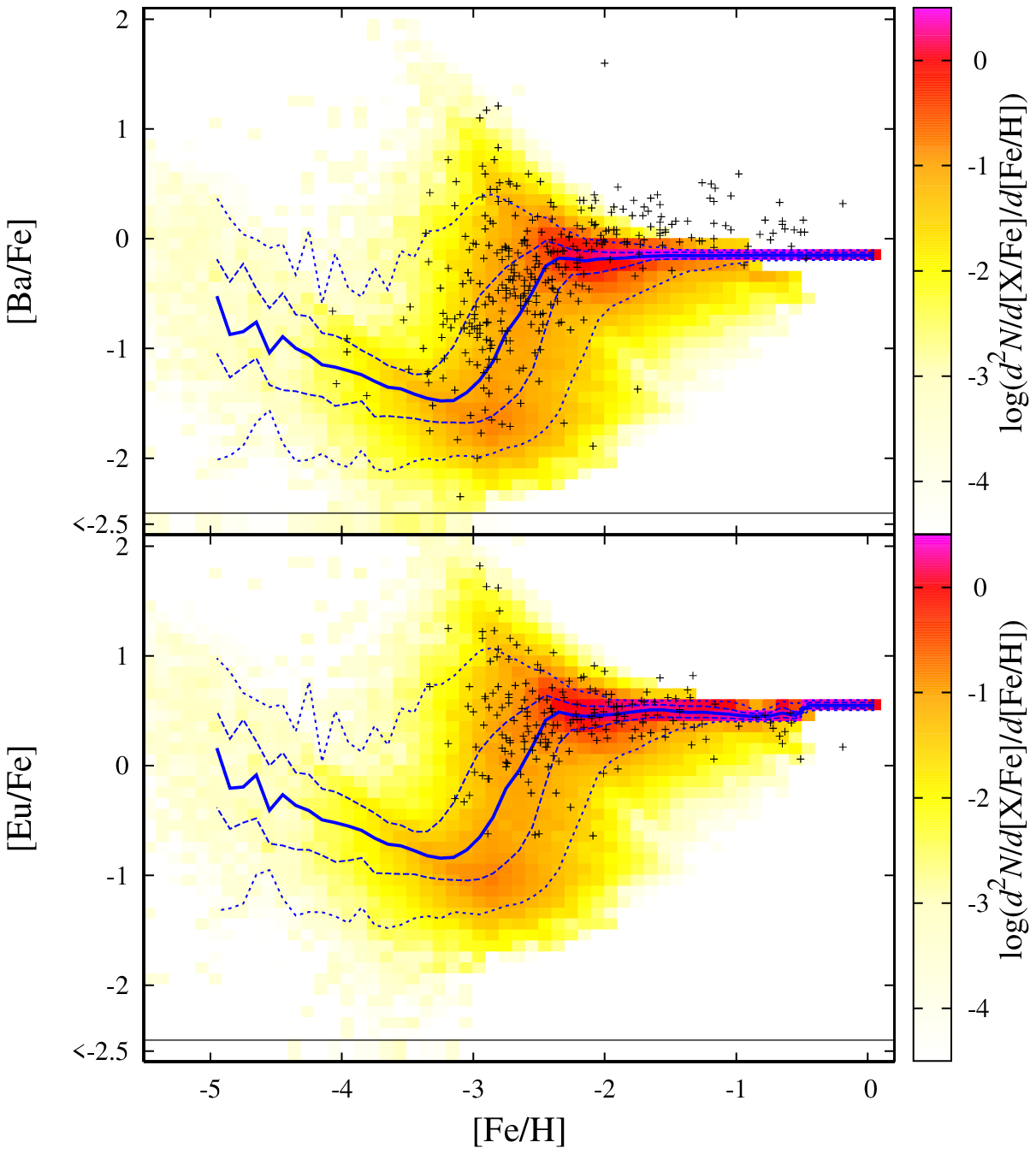}
\caption{
The predicted abundance distributions of stars on the $\feoh - \abra{Ba}{Fe}$ planes (top) and $\feoh - \abra{Eu}{Fe}$ planes (bottom) in the case of $f_{\rm esc} = 0$ (left) and $f_{\rm esc} = \exp(-R_{\rm g}/l_{\rm s})$ (right). 
The SFE depending on halo mass ($\propto M_{\rm h}^{0.3}$) is assumed. 
The band at the bottom edge of each panel shows the number of stars with $\abra{r}{Fe} < -2.5$ at each metallicity bin. 
The blue lines denote percentile curves of $5\%, 25\%, 50\%, 75\%$, and $95\%$ for the predicted distribution. 
The crosses show the observational data taken from the SAGA database. 
}\label{rFe}
\end{figure*}

\begin{figure*}[]
\includegraphics[width=0.5\textwidth]{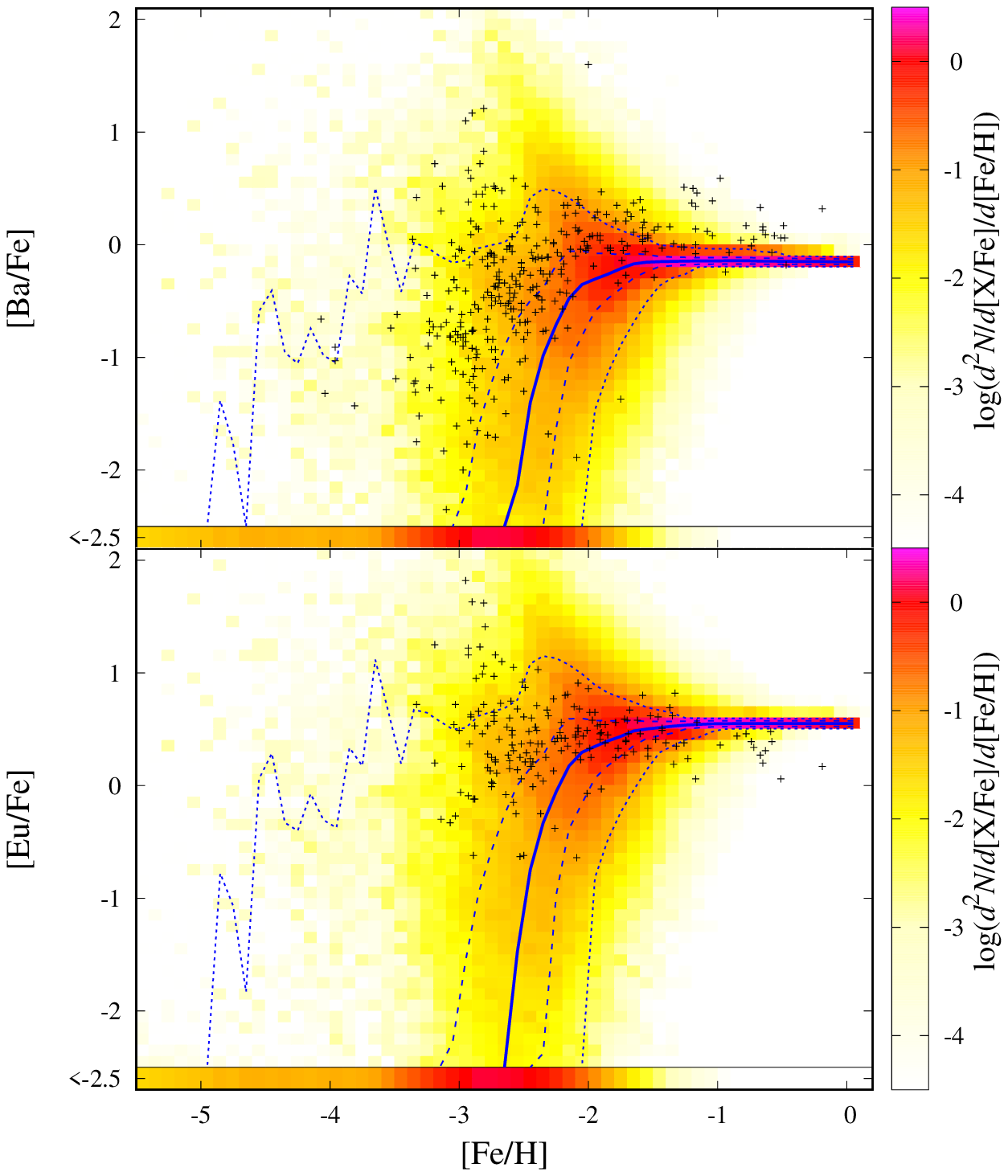}
\includegraphics[width=0.5\textwidth]{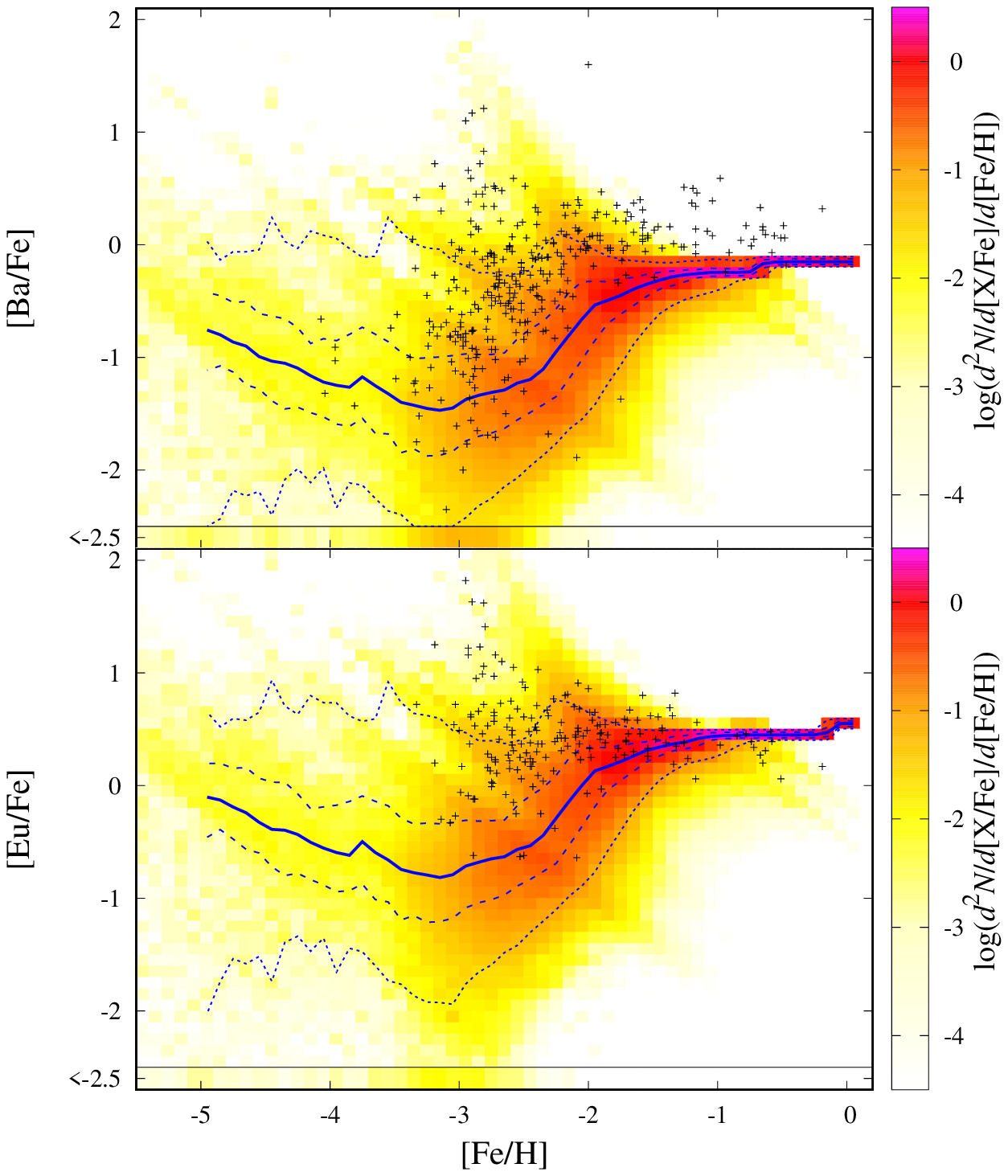}
\caption{
The same with Figure~\ref{rFe} but for model with constant SFE. 
}\label{outflow}
\end{figure*}

Figures~\ref{rFe} and \ref{outflow} show the r-process abundance of stars against metallicity for the fiducial model and the constant SFE model, respectively. 
The left and right panels denote results without escape of NSM ejecta ($f_{\rm esc}=0$) and with escape, respectively.

The constant SFE significantly delays the enrichment of r-process elements compared to iron and 
the predicted abundances of EMP stars do not match observational results, 
as mentioned in previous chemical evolution studies. 
The predicted abundances of r-process elements are significantly lower than observations at $\feoh \lesssim -2$, and r-II stars with $\abra{Eu}{Fe} > +1$ around $\feoh \sim -3$ is not reproduced. 
On the other hand, 
if the SFE is lower at lower mass proto-galaxies, 
r-process elements emerge at $\feoh \sim -3$, even with the long average delay time ($>10^9$ yr). 
This is because that enrichment timescale of iron is longer for low-mass proto-galaxies in which EMP stars were formed 
and some NSMs can take place within the formation epoch of EMP stars. 
These are consistent with the result of the one zone model by \citet{Ishimaru15}.

When we do not consider the escape of NSM ejecta from the host proto-galaxies, however, 
the majority of stars with $\feoh < -3$ show very low r-process abundances with [Ba/Fe] $<-2.5$, 
which is apparently at odds with observations (see the top left panel of Fig. 3). 
On the other hand, our escape model predicts $\abra{Ba}{Fe} \sim -1$ or $-2$ for the majority of stars at $\feoh \lesssim -3$, 
 and well reproduces observations of halo stars, as shown in the right top panel.

Proto-galaxies lacking NSM events are polluted by r-process elements ejected from NSMs in other proto-galaxies. 
As mentioned, IGM pre-enrichment is mostly responsible for the r-process elements in these stars, 
and capture of high-energy r-process elements also contribute at $z>10$. 
Stars in proto-galaxies without a NSM are distributed at $-6 \lesssim \abra{Ba}{H} \lesssim -4$. 
NSM events of short delay time, $t_d \lesssim 10^8 {\rm yr}$, are the dominant sources of r-process elements for these stars. 
On the other hand, proto-galaxies in which NSM events take place form stars with $\abra{Ba}{H} \gtrsim -3$.

Surface pollution can play an important role for stars in proto-galaxies without a NSM.  
Without surface pollution, there emerges a depression in the number density of stars at $\feoh \sim -3$ in the range of $-1 < \abra{Ba}{Fe} < 0$ where the first NSM occurs. 
The surface pollution alleviates the gap. 
In the model with $f_{\rm esc}=0$, the surfaces of most EMP stars have a tiny amount of Ba such as $\abra{Ba}{Fe} < -2.5$ even after surface pollution.

At $\feoh > -2$, the r-process abundance is averaged.  
The predicted Ba abundance is lower than observations since we neglect the s-process contribution.

At $\feoh \gtrsim -1$, while we can see a clear decreasing trend of $\abra{Mg}{Fe}$ as metallicity increases, 
we do not see a similar ``knee" in the predicted r-process abundance distributions. 
This is because the increase of the NSM rate due to the long delay time compensates the increasing production rate of iron enhanced by SNe Ia.
One possible explanation for the observed depression of $\abra{r}{Fe}$ is a lower frequency of NSMs with high metallicity progenitor stars, although we assume constant frequency in the computation for simplicity.

\begin{figure}[]
\includegraphics[width=\columnwidth]{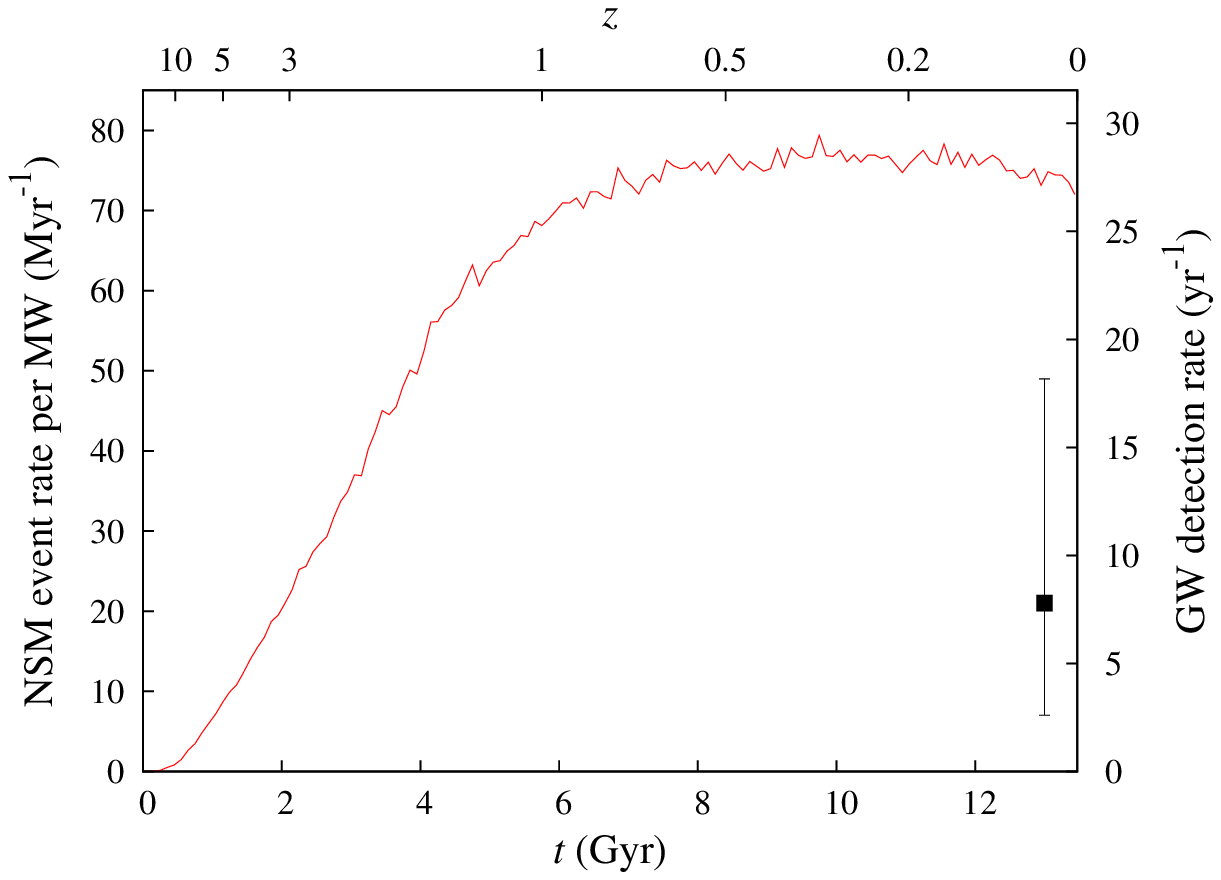}
\caption{
Event rate of NSMs in the computation box. 
The expected detection rate of gravitational wave (GW) by the advanced LIGO - Virgo network is presented at the right axis \citep{Abadie10}. 
A filled square refer to the estimation based on the observations of the binary pulsars in the MW \citep{Kim15}. 
}\label{NSMrate}
\end{figure}

The event rate of NSMs in this model is $\sim 60 \ {\rm Myr}^{-1}$ in the current MW, as shown in Figure~\ref{NSMrate}. 
It is also slightly higher than the estimation based on the observations of binary pulsars in the MW \citep[$21^{+28}_{-14} \ {\rm Myr}^{-1}$, ][]{Kim15}. 
Both the binary pulsars and the abundances of the metal rich stars indicate a factor of $\sim 3$ smaller NSM fraction than used in this paper at $z=0$. 
At the right axis of Figure~\ref{NSMrate}, the event rate in the MW is converted into the expected detection rate of the gravitational waves (GWs) by the advanced LIGO-Virgo network. 
Ongoing observations of GWs will provide better constraints on the NSM rate.

For the metallicity dependence of the NSM rate, studies of binary population synthesis predict a lower NS-NS merger rate for lower metallicity stars \citep{Dominik12, Kinugawa14}. 
For the BH-NS merger rate, on the other hand, \citet{Dominik12} predict that a rate for stars with the metallicity of $0.1\Zsun$ is higher than the Population I (Pop I) stars while \citet{Kinugawa14} predict that a rate for Pop III stars is lower than that for Pop I stars. 
Further studies are required for the event rate, delay time distribution, and r-process yield of NSMs, and their metallicity dependences.

In this paper, we assume that high-energy r-process elements go straight until they lose energies by interaction with gas particles. 
If proto-galaxies have magnetic field, however, particles can travel along a winding path, which decreases the escape probability and increases the capture rate. 
On the other hand, 
 the escape probability can be higher in a disk-like gas density profile than in a gas sphere assumed in this work. 
In addition, NSM events at the outer region of a proto-galaxy will effectively transfer r-process elements into intergalactic space, although we only consider NSM events at the center of proto-galaxies. 
Furthermore, a NS binary itself can go away from a proto-galaxy if it can get enough velocity from SN explosions. 
The velocity of a NS binary is depending on its binary orbit and SN kick. 
Although we still have no reliable estimation for the velocity distribution of NS binaries, two observed binary pulsars show a proper motion of more than $\sim 70$ km/s \citep{Beniamini16}, which exceeds the typical escape velocity of a proto-galaxy hosting EMP stars. 
The IGM pre-enrichment and transfer of r-process elements between proto-galaxies are depending on these details, but a model taking them into account is beyond the scope of this paper.

\section{Summary}\label{summaryS}
Coalescence of NS binaries is one of the most plausible sites for r-process nucleosynthesis but results of previous chemical evolution models with NSMs as major r-process sources is inconsistent with the observed r-process abundances of metal-poor stars.  
We revisit the Galactic chemical evolution of r-process elements following the NSM scenario.

The novelty of this paper is in considering the effect of propagation of NSM ejecta across proto-galaxies. 
As pointed out by \citet{Tsujimoto14}, r-process elements from NSMs have very large stopping lengths due to their high velocities. 
We compute the escape of r-process elements from proto-galaxies, pre-enrichment of IGM, and capture of the escaped elements by other proto-galaxies in the chemical evolution model based on the framework of the hierarchical galaxy formation. 

We have confirmed that the NSM scenario can reproduce the emergence of r-process elements at $\feoh \sim -3$
 in the case of lower SFE for less massive proto-galaxies, 
 while the constant SFE delays the enrichment of r-process elements, which results in abundance correlations incompatible with observations.

While previous chemical evolution models with the NSM scenario predict many stars lacking r-process elements at very low metallicities, 
 our model with the long-distance propagation of NSM ejecta predicts $-6 \lesssim \abra{Ba}{H} \lesssim -4$ for stars formed in proto-galaxies without a NSM. 
Our model succeeds in reproducing the r-process abundance distribution of EMP stars. 
This result indicates that the NSM scenario is not rejected by the Galactic chemical evolution, 
when we consider the propagation of NSM ejecta beyond the proto-galaxies.

Our current model predicts a slightly higher event rate of NSMs and results in $\abra{Eu}{Fe}$ higher than observations in the current MW while the good consistency with observations at very low metallicity.  
It may indicate the metallicity dependence of the NSM rate. 
Future observations of GWs and their electromagnetic counterparts will give us information about the NSM rate and its contribution to the Galactic chemical evolution.

\appendix

\section{Boundary Condition}
We adopt a periodic boundary condition which assumes that the high energy r-process elements incoming from outside the simulation box balance the elements escaping from the box. 
This seems to be a reasonable assumption at very high redshift since large scale density fluctuation tends to grow at low redshift. 
For example, at $z=20$, only $3 \sigma$ fluctuation of $M_{\rm h} <  10^6\ \msun$ collapses, and the box size is much larger. 
On the other hand, it is not a good approximation at low redshift since the environment around the current MW is significantly different from the average of the universe.

In the current universe, there are $0.0116 \ {\rm Mpc^{-3}}$ MW equivalent galaxies (MWEGs) on average \citep{Abadie10}, 
but the number density in our computation box corresponds to $0.0408 \ {\rm Mpc^{-3}}$ ($ = 1/V_{\rm MW}$). 
It may indicate that the production rate of high energy r-process elements (the first term of eq.~\ref{eq:HE}) may be overestimated by factor $ a = 0.0408/0.0116 \sim 3.52$ at $z=0$. 
The capture rate by proto-galaxies (the second term of eq.~\ref{eq:HE}) may also be overestimated since the capture rate in this computation box is dominated by the MW at low redshift.  
In order to mimic this difference, we compute the chemical evolution model with replacing the eq.~(\ref{eq:HE}) to the following equation, 
\begin{align} \label{eq:HEf}
\frac{dM_{i,{\rm HE}}}{dt} = 
\frac{1}{a} Y_i \sum_{n} \sum_{k} & \delta(t - t_{{\rm NSM}, n, k}) f_{{\rm esc}, n, i} \left( \sum_{m \neq n} \frac{\pi R_{{\rm g}, m}^2 (1-f_{{\rm cap}, m,i})}{4 \pi d_{n,m}^2}   \right) \notag \\
 &- \frac{1}{a} \sum_{n} \frac{M_{i,{\rm HE}} \pi R_{{\rm g}, n}^2 v_r}{V_{\rm MW}} f_{{\rm cap}, n,i}  - \frac{M_{i,{\rm HE}}}{\tau_{s,{\rm IGM}, i}}, 
\end{align} 
where $a = 3.52$ is the relative number density of MWEGs.

\begin{figure}[]
\includegraphics[width=\columnwidth]{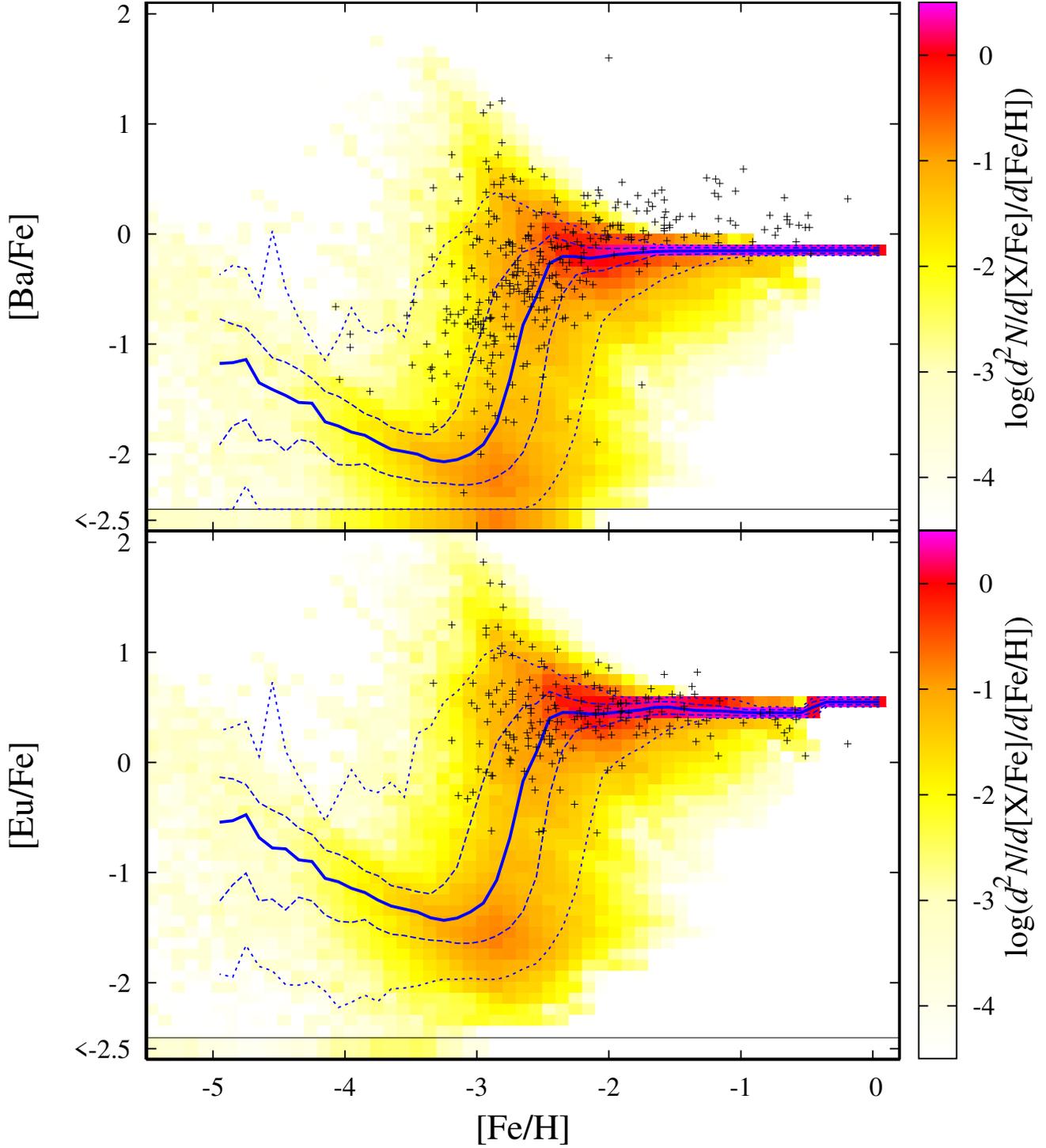}
\caption{
The same with the right panel of Figure~\ref{rFe} but for the different boundary condition. 
Incoming rate of high energy r-process particles from boundaries are smaller than outgoing rate by proportional to the relative density of Milky Way equivalent galaxies. 
See text for details. 
}\label{boundary}
\end{figure}

Figure~\ref{boundary} show the result. 
The predicted abundance distribution is indistinguishable from the fiducial model at high metallicity or high r-process abundance ($\abra{Ba}{H} \gtrsim -3$). 
This is because  
r-process elements of proto-galaxies in which these stars are formed are dominated by their own NSMs. 
On the other hand, the abundance of proto-galaxies without NSM events decreases since the r-process element abundances of the IGM decrease.  
The abundance of EMP stars at $\abra{Ba}{H} \lesssim -4$ formed before the first NSM in their host proto-galaxies decreases by $\sim 0.5$ dex. 
We note that the difference can be overestimated at low metallicity since the region around the MW should be more similar to the average of the universe at high redshift.

\end{document}